\documentclass[12pt,preprint]{aastex}

\def\hide#1{}

\begin{document}
\title{Coupling between the 45 Hz Horizontal-Branch Oscillation and the Normal Branch Oscillation in Scorpius X$-$1}
\author{Wenfei Yu\altaffilmark{1}}
\altaffiltext{1}{Shanghai Astronomical Observatory, 80 Nandan Road, Shanghai 200030, China. E-mail: wenfei@shao.ac.cn }
\begin{abstract}
The observations of the bright persistent neutron star low-mass X-ray binary (LMXB) Sco X-1 performed with the {\it Rossi X-ray Timing Explorer} (RXTE) show a $\sim$ 6 Hz normal-branch oscillation (NBO), a $\sim$ 45 Hz horizontal-branch oscillation (HBO), and twin kHz quasi-periodic oscillations (QPOs) on its normal branch simultaneously. We have found that the fractional amplitude of the HBO corresponding to the NBO phase of high flux is 1.1\%, while that of the NBO phase of low flux is undetectable, with a 3$\sigma$ upper limit of 0.4\%, implying that the HBO strength varies with the NBO phase in an opposite way to that of the lower kHz QPO previously found, and suggests that the condition for the generation of the HBO is met when the NBO flux is high. The 6 Hz NBO in Sco X-1 connects the 45 Hz HBO and the twin kHz QPO together, showing a unique picture indicating a coupling between the QPOs, which has never been observed in other neutron star LMXBs. We discuss the implications for current models of the 45 Hz HBO, the 6 Hz NBO, and the twin kHz QPOs. 
\end{abstract}


\section{Introduction}
The typical Fourier power spectra of neutron star low mass X-ray binaries (LMXBs) are composed of distinguishable components in the frequency range from millihertz to kilohertz, including noise components and quasi-periodic oscillations (QPOs) (see van der Klis 2000). The bright persistent neutron star LMXBs are divided into two types based on the pattern they trace in the X-ray color-color diagram and the shape of their Fourier power spectra; one is the brighter ``Z'' sources and the other is the fainter ``Atoll'' sources (Hasinger \& van der Klis, 1989). For the ``Z'' sources, along the Z track in the X-ray color-color diagram, the mass accretion rate is speculated to increase from the horizontal branch (HB) to the normal branch (NB), and finally, to the flaring branch (FB). Similarly for the ``Atoll'' sources, the mass accretion rate is thought to increase from the island branch to the banana branch. Simultaneous noise components and QPOs are observed in most of these branches, and the power spectra are usually characterized by some noise components and various QPOs, such as the 18-60 Hz horizontal branch oscillation (HBO) and the 5-20 Hz normal branch/flaring branch oscillation (N/FBO) in Z sources, and the kilohertz QPOs. 


The characteristic frequencies of most of the components are found to correlate with each other (e.g. Psaltis, Belloni \& van der Klis 1999; Belloni, Psaltis \& van der Klis 2002), indicating these frequencies are probably related to some characteristic time scales in the system, some of which may be used to constrain the mass and spin of the compact object. The frequencies of many QPOs in neutron star LMXBs, which are correlated with the X-ray count rate (e.g. Mend\'ez et al. 1998) and the X-ray spectral shape  (e.g. Kaaret et al. 1998), are thought to somehow represent the instantaneous mass accretion rate. These characteristic frequencies and their correlation are the main properties that current QPO models are aimed at to explain (e.g. the magnetospheric beat-frequency models for HBOs: Alpar \& Shaham 1985; Lamb et al. 1985; the sonic-point and spin-resonance frequency model for kHz QPOs: Miller, Lamb \& Psaltis 1998; Lamb \& Miller 2004; Relativistic precession model for HBOs and kHz QPOs: Stella \& Vietri 1998; Stella, Vietri \& Morsink 1999; Two-oscillator model for most of the QPOs now known in NS XRBs: Titarchuk \& Osherovich 1999; Titarchuk, Osherovich \& Kuznetsov 1999; the non-linear resonance model: Abramowicz \& Kluzniak 2001; the hot-spot model: Schnittman \& Bertschinger 2004; Alfve\'n wave oscillation model: Zhang 2004; Kink mode model: Li \& Zhang 2005). However, in most of the modeling, the generation of the X-ray flux of the QPO signal is missing. In the real situation, even if a certain mode of high coherence in the accretion flow exists, it may not be able to propagate to the observer as an oscillation in the X-ray flux. Observations show that the spectra of most of these variability components are harder than the persistent spectra, showing an increasing fractional amplitude towards higher photon energies in the RXTE/PCA energy band. Mechanisms to generate the hard X-ray flux oscillating at the characteristic frequencies are needed for most of the models. Phase resolved spectral study of the QPOs and identification of QPOs as  amplitude modulations are potentially useful to determine the origin of the QPO flux. 

Sco X-1 is the brightest neutron star low-mass X-ray binary. Its high X-ray flux makes it one of the best targets among the neutron star LMXBs for studying the relation between individual timing features and their coupling. Its RXTE/PCA count rate is so high that other ``Z'' sources can only be detected at the same count rate with an instrument of about ten times the area of  the RXTE/PCA.  If we ignore the dead-time effect, which is a disadvantage of the RXTE observations of Sco X-1, the RXTE observations of Sco X-1 would be comparable to future observations of other ``Z'' sources with a next generation X-ray timing mission. 

In the RXTE observations of its normal branch, Sco X-1 has shown a QPO feature at 45 Hz with a harmonic component at about 90 Hz, which is identified as the HBO and its first harmonic. This QPO is simultaneously observed with the 6--8 Hz normal branch oscillation and the twin kHz QPOs (van der Klis et al. 1997). The study of the kHz QPOs in relation to the NBO phase shows that the upper kHz QPO frequency is anti-correlated with the NBO flux. The upper kHz QPO frequency on average is about 22 Hz higher during the phase of low NBO flux than that during the phase of high NBO flux. At the same time, the lower kHz QPO appears stronger during the phase of low NBO flux. The anti-correlation between the upper kHz QPO frequency and the NBO flux is explained as that the NBO flux is generated inside the inner disk edge, and the radiative force of the NBO flux on the accretion flow slows down the orbital frequency in the accretion (Yu, van der Klis \& Jonker 2001). The dependence of the kHz QPO properties on the phase of the NBO makes the NBO a unique phenomena, as it connects several major variability components in neutron star LMXBs. 

In this Letter, we show that the amplitude of the 45 Hz HBO in the ``Z'' source Sco X-1 varies with the NBO flux in a way opposite to that of the lower kHz QPO (Yu, van der Klis, \& Jonker 2001). Especially, the 45 Hz HBO disappears during the NBO phase of low flux. The variation of the HBO amplitude with the phase of the NBO is consistent with the idea  that the condition for the generation of the HBO is met in Sco X-1 during the NBO phase of high flux. 

\section{Observations}
The observations of Sco X-1 performed with the {\it Rossi} X-ray Timing Explorer (RXTE) on  the morning of Jan 28, 1996 were used in our analysis. The observation IDs are 10056-01-04-01 and 10056-01-04-02. The high time resolution data mode {\it B\_125us\_1M\_0\_87\_H}, which covers the PCA original energy channel range 0--87, is used to generate the light curves. In most of the segments of the observations (segment 1,2 and 4 of the 4 segments in the observation 10056-01-04-01, and segment 2,3,4 of the 4 segments in the observation 10056-01-04-02), both the $\sim$ 6 Hz NBO and the $\sim$ 45 Hz QPO were significant and well-constrained in frequency, with an rms of 2.5\% and 0.8\%, respectively. The average count rate is about 80000 c/s. We have only selected observations in those analyzed in Yu, van der Klis \& Jonker (2001) when the NBO and the 45 Hz QPO were strong. For the individual segment in which there was a change of the number of proportional counter unit (PCU) on board the RXTE, we have chosen the longer part of the segment corresponding to the same number of PCU. The typical average power spectrum of Sco X-1 corresponding to these observations is shown in Fig.~1. It shows that VLFN, $\sim$ 6 Hz NBO, $\sim$ 45 Hz HBO and the kHz QPOs coexist.  

\section{Analysis and Results}
Since we have chosen the observations where the NBO was very strong and narrow in the Fourier power spectra, the NBO waveform in these observations were mostly visible in the X-ray light curve. We first determine those intervals corresponding to the high and the low fluxes of each NBO oscillation cycle in the X-ray light curve as follows. We have identified the local maxima and the local minima of the X-ray flux on the time scale of 0.125 seconds within each time window of 0.5 second, with a time accuracy of the identifications being 0.03125 seconds. Then we select those intervals corresponding to the local maxima and the local minima where they are next to each other within 1.0 second, so that they roughly corresponds to the same NBO cycle (as shown in Fig.~1, the lower frequency boundary of the NBO is about 1 Hz). In total we have obtained about 2400 pairs of the intervals corresponding the NBO phases of high flux and low flux, respectively. These adds up to about 1/9 of the total observation time. 

We have calculated the Fourier power spectra corresponding to both groups of intervals. We use the data combined from the two data modes of high time resolution, and re-binned the light curve to a resolution of 1/4096~s. We have used transform lengths of 0.125~s, which have resulted 8~Hz resolution power spectra with a Nyquist frequency of 2048~Hz. The power spectra corresponding to the NBO phase of high flux and the NBO phase of low flux have been averaged, and are shown in Fig.~2 as filled circles and crosses, respectively. The amplitude of the 45 Hz QPO in the average power spectrum corresponding to the NBO phase of high flux is very strong, with a fractional rms amplitude of 1.1\%, in contrast to that corresponding to the NBO phase of low flux which is not detected, with a 3 $\sigma$ upper limit of 0.4\%. The residual power in the frequency range of the 45 Hz QPO and its second harmonic is also shown in the inset panel of Fig.2. The power difference in the 45 Hz range is very significant. This was also seen in the study of the relation between the NBO and the twin kHz QPOs using a different method (Yu, van der Klis \& Jonker 2001). In the current study we have tried to use a shorter Fourier transform (i.e., 0.0625 s) to study the 45 Hz HBO strength. As the smaller transform length yields a larger frequency resolution, i,e,. 16~Hz, and thus sacrifices the sensitivity of analysis for the HBO detection, we could not improve our results by decreasing the transform length further. There is a feature around 65 Hz of low significance in the overall average power spectra as well as in the average power spectra corresponding to the NBO phase of low flux. We can not determine if it is the HBO shifted in frequency with the NBO phase. A future detection of a QPO at 65 Hz as the shifted HBO would be crucial to some QPO models. This will be discussed in the next section. 

\section{Discussion and Conclusion}
By analyzing RXTE observations of Sco X-1 when the $\sim$ 6 Hz NBO and the $\sim$ 45 Hz HBO were simultaneously detected, we have found that the amplitude of the 45 Hz HBO strongly depends on the NBO phase. During the NBO phase of low flux, the HBO was not detected, while during the NBO phase of high flux, the HBO is significant. This indicates that the HBO appears coupled with the NBO phase, and the result connects the two QPOs in the neutron star LMXBs. Because the fractional {\it rms} amplitudes of the NBO and the HBO are 2.5\% and 0.8\%, respectively, the amplitude of the NBO is more than twice that of the HBO. Thus it is possible that the 45 Hz HBO is a modulation of the NBO flux. 

The coupling between the HBO amplitude and the NBO flux may be generated in two ways. One is that the HBO, which is usually generated under certain accretion conditions on the HB, occurs when the same conditions are met on the NB at the NBO phase corresponding to a high NBO flux. The observations of the HBO and the NBO  in Sco X-1 show that they co-exist on part of the NB. Such a co-existence is not typical in the ``Z'' sources. We speculate that in Sco X-1, when the NBO occurs on the NB and the NBO flux increases to a certain level, the condition for the generation of the HBO is met. When the NBO flux decreases below that level, the condition is not met and the generation of the HBO is terminated. As the NBO flux is suspected to originate from inside the inner disk edge which might lead to significant movement of the inner disk radius in Sco X-1 (Yu, van der Klis and Jonker 2001), the generation of the 45 Hz HBO at a higher NBO flux may be understood as that the radius of the inner disk edge has moved outward so that the accretion configuration is similar to that on the HB when the HBOs are usually generated. Alternatively, the disappearance of the HBO may be associated with the occurrence of the radial inflow that might exist throughout the NB, which is also responsible for the NBO as proposed in the NBO model by Fortner, Lamb, \& Miller (1989). When at the NBO phase of low flux, the radial inflow may be temporarily halted, and the HBO disappears. The other possibility is that the HBO and the NBO are directly related in flux. For example, the HBO photons may be part of the NBO photons, as  the NBO photon flux is modulated at 45 Hz and seen as the HBO. In this case, the rms amplitude of the 45 Hz HBO should be always smaller than the rms amplitude of the NBO. This is consistent with the measurement of the HBO and the NBO amplitudes, and can be tested in future observations. In this case the photons producing the HBO and the NBO signals should originate from the same site. Because the HBO usually appears on the HB when there is no NBO, it is reasonable to speculate that the HBO is a modulation of the photon flux of the same source of the NBO. To summarize, the coupling between the NBO and the 45 Hz HBO gives us important information about the nature of these QPOs and the difference between the HB and the NB, especially about the conditions for the generation of the HBO. 

We can compare our results with the predictions of the models involving HBOs. The NBO flux is anti-correlated with the upper kHz QPO frequency and the amplitude of the lower kHz QPO in Sco X-1 (Yu, van der Klis, \& Jonker 2001). Comparing the HBO with the lower kHz QPO, the relation of the HBO amplitude to the NBO flux is opposite to that of the lower kHz QPO to the NBO flux. This seems to contradict the relativistic precession model (Stella \& Vietri 1998; Stella, Vietri \& Morsink 1999) which suggests that the HBO and the kHz QPOs are related to the same slightly tilted eccentric orbits in the innermost disk region. As it was originally proposed, the HBOs and the upper kHz QPOs are both generated from a ring of gas at a special radius in the inner disk. The lower kHz QPOs come from the relativistic periastron precession of the eccentric ring and the HBOs come from the nodal precession of the same region. Thus, this model favors a positive correlation between the HBO strength and the low frequency kHz QPO strength. On the other hand, the HBO frequency and the upper kHz QPO frequency should follow a positive correlation which should not vary. The HBO frequency has been observed relatively constant at around 45 Hz and the upper kHz QPO varies more than 22 Hz in an NBO cycle (see Yu, van der Klis, \& Jonker 2001). In order to explain the difference, the HBO strength should be weak enough during the NBO phase of low flux ($>$ 0) so that we are not able to detect the variation. We have noticed that there might be some excess in Fourier power in the frequency range 60--70 Hz in the average power spectra corresponding to the NBO phase of low flux (Fig.~2) and in the overall average power spectra. Determination of the association of the 60--70 Hz feature with the 45 Hz HBO would clarify if the relativistic precession model is consistent with the observation of the frequency movements. A similar argument also applies to other models (e.g., Zhang 2004).

In the Two-Oscillator model (Titarchuk \& Osherovich 1999; Titarchuk, Osherovich \& Kuznetsov 1999), the lower kHz QPO frequency is the Keplerian frequency, while the upper kHz QPO is the radial mode of a Keplerian oscillator associated with a hot blob thrown into the magnetosphere. The HBO is the corresponding perpendicular mode of the same Keplerian oscillator. Thus the upper kHz QPO, the lower kHz QPO, and the HBO are associated with the same hot blob. Similarly to the discussion of the relativistic model about the variation of the HBO amplitude, which is opposite to that of the lower kHz QPO, the model is not able to explain why the amplitude of the HBO and that of the lower kHz QPO show the opposite response to the NBO flux. Regarding the frequency shift, in this model, the HBO frequency ($\nu_{L}$) is proportional to $\nu_{k}$/$\nu_{h}$ (see formula (2) in Titarchuk, Osherovich \& Kuznetsov 1999). So, the HBO frequency will not shift significantly (only $\sim$ 2\% -- 0.1 Hz) when the upper kHz QPO frequency ($\nu_{h}$) has a shift of 22 Hz ($\sim$ 2\% of $\nu_{2}$). If taking into account the NBO radiation force from inside the inner disk edge which acts in the same way as the Coriolis force, an additional factor on $\Omega$ is needed. This effect leads to an opposite slight movement of the HBO frequency to the original frequency. So, the HBO frequency is about constant regarding a 22 Hz upper kHz QPO frequency shift in the TO model. This qualitatively agrees with our current results. 

Regarding the sonic point and spin resonance model for kHz QPOs (Miller, Lamb \& Psaltis 1998; Lamb \& Miller 2002 ) and the magnetosphere beat-frequency model for the HBOs (e.g. Alpar \& Shaham 1985; Lamb et al. 1985; Psaltis et al. 1999), the kHz QPO frequency and the HBO frequency are generated from distinct radii, i.e. at the sonic point and the magnetospheric boundary, respectively. This model may explain the occurrence of the HBO as that the condition at the magnetospheric boundary is similar to that on the HB during the NBO phase of high flux. This may be because of the effect of the radiation force of the NBO on the inner disk radius, or because of a temporary halting of the radial inflow. In these models, the frequency correlation should be indirect, since the correlation is originated from the frequency correlation with the local mass accretion rate. Psaltis et al. (1999) have calculated the correlation between the HBO frequency and the upper kHz QPO frequency in the framework of beat-frequency models. Following their eq. (5) and taking the parameters in their Table.2 for Sco X-1, we get a HBO frequency variation of $\sim$ 0.7 Hz if there is a 22 Hz shift of the upper kHz QPO. We can not detect such a frequency shift from the current observations. 

The twin kHz QPOs might be the same phenomena as the high frequency QPOs in black hole binaries, which shows a 3:2 frequency ratio  (e.g., Abramowicz et al. 2004). The non-linear resonance model (Abramowicz \& Kluzniak, 2001; Abramowicz 2005) explains the twin high QPO frequencies as a non-linear resonance between two weakly coupled modes of accretion disk oscillations. It also explains that some low frequencies appear. A possible explanation of the NBO 6 Hz frequency in Sco X-1 was already discussed in detail by Hor\'ak et al. (2004), as a modulation due to the energy exchange between the twin high frequency modes. It was also suggested by Abramowicz et al. (2004) although not explained in details, that a second low frequency should appear due to effect known in oceanography as the 9th wave (the Benjamin-Frei instability). On the other hand, the observations discussed here imply that these QPOs are of an oscillatory nature, with several high and low frequency oscillations being coupled in frequency as well as in amplitude, which is inconsistent with the interpretation adapted from the hot-spot interpretation of the high frequency QPOs in black hole systems (e.g. Schnittman \& Bertschinger E., 2004; Schnittman, 2005).

Although further work on the coupling between the QPOs in neutron star and black hole X-ray binaries are needed to address their nature, the discovery of the dependence of the HBO strength on the NBO phase confirms that the X-ray variability of the X-ray binaries, composed of noise components and QPOs, could vary drastically on short time scales in both frequency and amplitude. The measurements of the frequency of QPO and noise components from average power spectra should be used with caution. Especially for Sco X-1, we have found both frequency shift and amplitude variation on time scales of 0.08 s or shorter (Yu, van der Klis, \& Jonker 2001). This would significantly affect the measurement of the frequency separation between the two kHz QPOs -- one of the distinguishable predictions of current models of kHz QPOs (see van der Klis 2000). 

\acknowledgments
WY would like to thank the anonymous referee for useful insight into current QPO models. WY would like to thank Prof. Michiel van der Klis for very significant contribution to this and related research as well as continuous encouragement, who deserves to be a coauthor of this paper. WY would also like to thank the Astronomical Institute of the University of Amsterdam where the idea of this paper originated. This work was supported by the {\it One Hundred Talents} Project of the Chinese Academy of Sciences initiated at SHAO, and has made use of data obtained through the High Energy Astrophysics Science Archive Research Center Online Service, provided by the NASA/Goddard Space Flight Center.

\newpage
\begin{figure}
\plotone{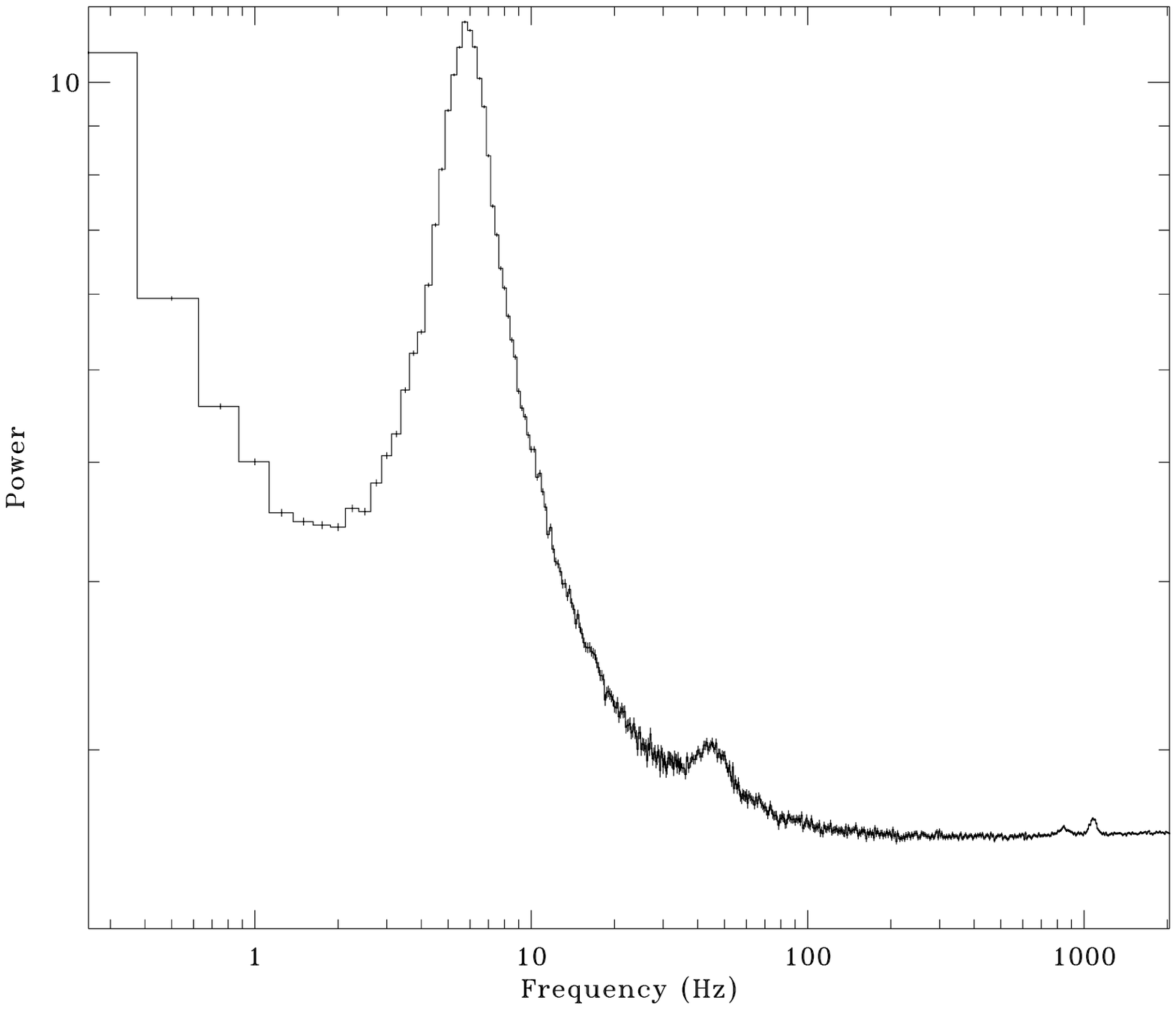}
\caption{The average power spectrum of the X-ray light curve of Sco X-1 in the 
RXTE observations we analyzed. It shows the strong NBO around 6 Hz, the HBO around 45 Hz, and the twin  kHz QPOs in the 800 -- 1100 Hz range. }
\end{figure}

\begin{figure}
\plotone{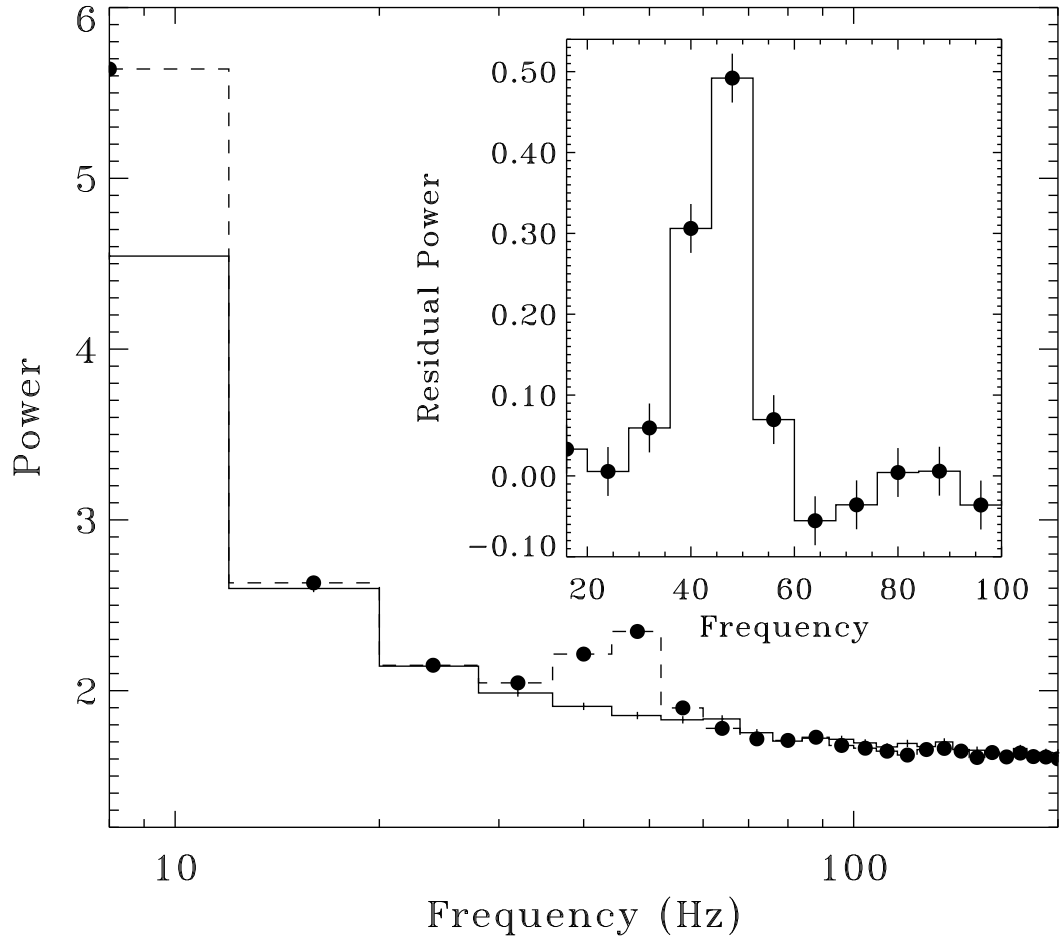}
\caption{Average power spectra corresponding to the NBO phases of high flux (filled circles) and low flux (crosses), showing significant difference in the 45 Hz HBO range. The residual power between the two is shown in the inset panel. The significance of the power difference around 45 Hz is more than 10 $\sigma$. Notice that the 45 Hz HBO is consistent with a non-detection during the NBO phase of low flux.}
\end{figure}


\begin{thebibliography}{}

\bibitem[Abramowicz01]{abramowicz01} Abramowicz, M. A. \& Kluzniak, W. 2001, A\&A, 374, L19

\bibitem[Abramowicz04]{abramowicz04} Abramowicz, M. A., Kluzniak, W. Stuchlik, Z. et al. 2004, Astro-ph (0401464)

\bibitem[Abramowicz05]{abramowicz05} Abramowicz, M. A. 2005, Astronomische Nachrichten, Vol. 326, Issue 9, 782 

\bibitem[Alpar \& Shaham 1985]{alpar85} Alpar, M. A., \& Shaham, J. 1985, Nature, 316, 239 

\bibitem[Belloni et al.(2002)]{2002ApJ...572..392B} Belloni, T., Psaltis, 
D., \& van der Klis, M.\ 2002, \apj, 572, 392

\bibitem[Kaaret et al.(1998)]{1998ApJ...497L..93K} Kaaret, P., Yu, W., 
Ford, E.~C., \& Zhang, S.~N.\ 1998, \apjl, 497, L93 

\bibitem[Horak et al. 2004]{horak04} Hor\'ak, J., Abramowicz, M. A., Karas, V. et al. 2004, PASJ, Vol. 56, No. 5, 819

\bibitem[Lamb et al. 1985]{lamb85} Lamb, F. K., Shibazaki, N., Alpar, M. A., \& Shaham, J. 1985, Nature, 317, 681

\bibitem[Li \& Zhang 2005]{li05} Li, X.-D.\& Zhang, C. M. 2005, ApJ, 635, L57

\bibitem[M{\'e}ndez et al.(1999)]{1999ApJ...511L..49M} M{\'e}ndez, M., van 
der Klis, M., Ford, E.~C., Wijnands, R., \& van Paradijs, J.\ 1999, \apjl, 
511, L49 

\bibitem[Miller, Lamb \& Psaltis 1998]{miller98} Miller, C. M., Lamb, F. K.
\& Psaltis, D. 1998, ApJ, 508, 791

\bibitem[Psaltis et al.(1999)]{1999ApJ...520..262P} Psaltis, D., Belloni, 
T., \& van der Klis, M.\ 1999, \apj, 520, 262 

\bibitem[Schnittman \& Bertschinger 2004] Schnittman, J. D. \& Bertschinger, E. 2004, ApJ, 606, 1098

\bibitem[Schnittman 2005] Schnittman, J. D. 2005, ApJ, 621, 940

\bibitem[Stella \& Vietri(1999)]{1999PhRvL..82...17S} Stella, L., \& 
Vietri, M.\ 1999, Physical Review Letters, 82, 17 

\bibitem[Stella et al.(1999)]{1999ApJ...524L..63S} Stella, L., Vietri, M., 
\& Morsink, S.~M.\ 1999, \apjl, 524, L63 

\bibitem[Titarchuk \& Osherovich(1999)]{1999ApJ...518L..95T} Titarchuk, L., 
\& Osherovich, V.\ 1999, \apjl, 518, L95 

\bibitem[Titarchuk et al.(1999)]{1999ApJ...525L.129T} Titarchuk, L., 
Osherovich, V., \& Kuznetsov, S.\ 1999, \apjl, 525, L129 

\bibitem[van der Klis 2000]{klis00} van der Klis, M. 2000, ARA\&A,  38, 717

\bibitem[van Straaten et al.(2003)]{2003ApJ...596.1155V} van Straaten, S., van der Klis, M., \& M{\'e}ndez, M.\ 2003, \apj, 596, 1155

\bibitem[Yu, van der Klis \& Jonker 2001]{yu01} Yu, W. \& van der Klis, M., Jonker, P. 2001, \apj, 559, L29

\bibitem[Zhang 2004]{zhang04} Zhang, C. M. 2004, A\&A, 423, 401

\end{thebibliography}
\end{document}